\magnification 1200
\centerline { \bf $BVRI$ CCD photometric standards in the field of GRB 990123 }
\bigskip
\centerline {\it Nilakshi, R.K.S. Yadav, V. Mohan, A.K. Pandey and R. Sagar }
\medskip
\centerline {\it U. P. State Observatory, Manora Peak, Nainital -- 263 129,
 India}
\bigskip
\centerline {\bf Abstract}
\bigskip

The CCD magnitudes in Johnson $BV$ and Cousins $RI$ photometric passbands 
are determined for 18 stars in the field of GRB 990123. These measurements can
be used in carrying out precise CCD photometry of the optical transient of 
GRB 990123 using differential photometric techniques during non--photometric
sky conditions. A comparison with previous photometry indicates that the
present photmetry is more precise.
\bigskip

\bigskip

\centerline {\bf 1. Introduction}
\bigskip

The optical follow up observations of the transients of gamma-ray bursts (GRB) 
are valuable for understanding the nature of these bursts. Such observations
are therefore generally carried out even during non--photometric sky conditions.
The optical transient (OT) are generally faint and located in such regions of 
the sky where accurate photometric standards are generally not available 
beforehand. For reliable determination of the photometric magnitudes of the OT, 
such standards are desired. In this paper, we provide Johnson $BV$ and Cousins 
$RI$ magnitude of stars in the field of GRB 990123, first reported by 
Piro (1999). These observations can be used to
perform accurate photometry of the OT of GRB 990123, as most of its CCD 
observations have been carried out in non--photometric sky conditions.
Use of differential photometric technique can provide  accurate photometric
magnitudes of the OT from such observations if accurate photometric standards 
are present in the field on the CCD images of the OT. To provide photometric
standard stars in the field of GRB 990123, we observed the field along with
M67. A total of 18 stars in the field have been calibrated and their standard
$BVRI$ magnitudes are prsented here. The observations and data reduction have
been described in sections 2 and 3 respectively. A comparison of the present
photometry with earlier BR photometric measurements is given in section 4.

\bigskip

\centerline {\bf 2. Observations} 
\bigskip

The observations of the field containing GRB 990123 have been carried out from
23rd January to 25th January, 1999 using CCD system attached to the 104-cm
Sampurnanand telescope of the U.P. State Observatory, Nainital. The CCD camera
has a 1024 x 1024 pixel$^{2}$ Tektronix front illuminated detector. For present
observations
we used 2 x 2 binning on the CCD chip. Thus each superpixel of the CCD 
corresponds
to 0.76 arc second at the f/13 Cassegrain focus of the telescope. For 
calibrating the magnitudes, on the night of 25/26 January, 1999 the M67 
``dipper asterism" field was also observed. The observing conditions were 
photometric. The log of CCD observations is given in Table 1. In addition to
these observations, several twilight flat field and bias frames were also 
observed.

\bigskip

\centerline {\bf 3. Data Reduction} 
\bigskip

The CCD frames were cleaned using standard procedures. Image processing was done
using ESO MIDAS and DAOPHOT softwares. Atmospheric 
extinction coefficients were determined from the observations of the brightest
star present in the GRB field and these are used in further analysis.
Standard magnitudes of stars in the field of M67 were taken from
Chevalier and Ilovaisky (1991). A total of 29 stars were used to determine
the transformation coefficients and photometric zero points.They cover a wide 
range in 
colour ($-0.1 < B-V < 1.4$) as well as in brightness ($ 9.7 < V < 14.3 $). 
The transformation
coefficients were determined by fitting least square linear regressions to the
standard $BVRI$ photometric indices as function of the observed instrumental
magnitudes normalised for 1 second exposure time.  The
following colour equations were obtained for the system.
\bigskip

	$ B-V = (1.096 \pm .018) (b-v)_{CCD} - (0.763 \pm .023) $
\medskip
	$ V-R = (1.051 \pm .023) (v-r)_{CCD} - (0.383 \pm .017) $
\medskip
	$ V-I = (0.965 \pm .019) (v-i)_{CCD} - (0.123 \pm .017) $
\medskip
	$ V   = v_{CCD} + (0.015 \pm .013) (B-V) - (4.938 \pm .009) $
\bigskip

\noindent where $v_{CCD}, (b-v)_{CCD}, (v-r)_{CCD} $ and $(v-i)_{CCD}$
represent the instrumental colour indices corrected for atmospheric extinction.
The errors in the colour coefficients and zero points are obtained
from the deviation of data points from the linear relation.

In order to calibrate stars in the GRB field, firstly the magnitudes of
the brightest star (star no. 1 of Table 2) in the field were calculated using 
CCD frames observed on the night of 25/26 January, 99 and applying the above 
transformations. For increasing the photometric precision of fainter stars, 
$B,V$ and $R$ CCD images taken on the nights of 24/25 and 25/26 January, 99 were 
co-added in the same filter. Thus the total exposure in $B,V$ and $R$ filters 
becomes 80, 80 and 100 minutes respectively. $I$ magnitudes were determined 
using the observations of 25/26 January, 99 alone having a total exposure of 
10 minutes. From these added images, PSF magnitudes were determined using 
DAOPHOT software. The standard magnitudes of the stars were then determined 
differentially using the brightest star as the local standard star. The 
magnitudes thus determined along with their errors
are given in Table 2. In this table column 1 is the identification number; 
column 2 and 3 are the Right Ascension and Declination (Epoch 2000) of the 
stars in the CCD frame taken from the catalogue of Monet (1997); Column 4,
6, 8 and 10 give standard $V, (B-V), (V-R)$ and $(R-I)$ magnitudes 
respectively; and in
columns 5,7,9 and 11 are given the associated DAOPHOT errors. These errors are
 primarily related to signal to 
noise ratio and do not include the errors related to colour transformations
and magnitude zero points.

\vfill\eject

\centerline {\bf 4. Comparison with previous photometry}

\bigskip

In this section, we compare the present CCD photometry with available previous
$B$ and $R$ photographic magnitudes given in the PMM USNO -- A1.0 catalogue by
Monet (1997).  In figure 1, we plot the differences (prsent -- Monet) as a 
function of present $V$ magnitude and $(B-R)$ colour. There is a large 
difference in $B$ and $R$ magnitudes of four objects, out of which two 
objects (no. 7 and 9) are
galaxies. For the remaining 14 stars, the difference in $R$ and $B$ magnitudes
is $0.03 \pm 0.14$ and $-0.06 \pm 0.18$ respectively. Considering the fact that 
the $B$ and $R$ magnitudes of Monet (1997) are based on Palomar Schmidt plates,
the agreement between the two photometries is excellent.
The results of statistical comparison are given in Table 3. From this table 
and figure 1, it appears that these differences have dependence on 
brightness as well as on colour and have expectedly large scatter. 
Least square linear regression to data points yield following relations :
\bigskip

$ \Delta B = (0.004 \pm 0.028) V - (0.13 \pm 0.51) $
\medskip

$ \Delta R = (0.025 \pm 0.021) V - (0.41 \pm 0.38) $
\medskip
$ \Delta (B-R) = (-0.02 \pm 0.04) V + (0.26 \pm 0.65) $
\medskip
$ \Delta (B-R) = (-0.124 \pm 0.10) (B-R) + (0.06 \pm 0.15) $

\bigskip
Sokolov et al. (1999), reported that when estimating the magnitude of GRB 
990123, by using the $R$ magnitudes of star nos. 1 and 2 as references, there 
is a differnce of 0.3 magnitudes. However, if $R$ magnitudes of these reference
stars as determined by us are used, this discrepency vanishes. This 
indicates that the magnitudes obtained by us are more precise.

\bigskip

\centerline {\bf 5. Conclusion}

\bigskip

	We have determined $BVRI$ magnitudes of 18 stars in the field of 
GRB 990123. These magnitudes can be used to calibrate photometric magnitudes
of the OT of GRB 990123 even in non photometric sky conditions using 
differential photometric techniques. We have used these magnitudes to 
calibrate various photometric observations of the OT of GRB 990123 given in
GCN circulars. Based on these calibrated magnitudes, Sagar et al. (1999) have 
presented the $BVR$ light curves of the OT. A comparison with earlier photometry
indicates that present photometric magnitudes are reliable. 

\bigskip

\centerline {\bf References}
\bigskip

\item {} Chevalier, C., Ilovaisky, S.A., 1991, Astron. Astrophys. Suppl.,
                {\bf 90}, 225.
\item {} Piro., Luigi, 1999, GCN Observational Report No. {\bf 199}.
\item {} Monet, D., 1997, The PMM USNO-A1.0 Catalogue.
\item {} Sagar, R., Pandey, A.K., Mohan, V., Yadav, R.K.S., Nilakshi, 1999,
     Bull. Astron. Soc. India {\bf 27}, (accepted). 
\item {} Sokolov, V., Zharikov, S., Nicastro, L., Feroci, M., Palazzi, E., 
1999, GCN Observational Report No. {\bf 209}.
\vfill\eject

\noindent {\bf Table 1.} Log of Observations of GRB 990123.
\bigskip
\settabs \+ January 24/25 19 & fieldsssssss & filterfilss & exposure \cr
\hrule
\medskip
\+ Date & Field  & Filter & Exposure (in seconds) \cr
\medskip
\hrule
\medskip
\+ 24/25 Jan 99 & GRB & R & 1200 x 3 \cr
\+ 24/25 Jan 99 & GRB & B & 1200 x 3 \cr
\+ 24/25 Jan 99 & GRB & V & 1200 x 1 \cr
\+ 25/26 Jan 99 & GRB & B & 600 x 2 \cr
\+ 25/26 Jan 99 & GRB & V & 1200 x 3, 300 x 1 \cr
\+ 25/26 Jan 99 & GRB & R & 1200 x 2, 300 x 1 \cr
\+ 25/26 Jan 99 & GRB & I & 300 x 2 \cr
\+ 25/26 Jan 99 & M67 & I & 10 x 2, 8 x 1 \cr
\+ 25/26 Jan 99 & M67 & R & 8 x 1, 10 x 2, 15 x 1 \cr
\+ 25/26 Jan 99 & M67 & V & 60x 1, 40 x 1 \cr
\+ 25/26 Jan 99 & M67 & B & 120 x 1 \cr
\medskip
\hrule
\bigskip

\noindent {\bf Table 2.} $BVRI$ Standard magnitudes of the objects in the region 
of GRB 990123.

\bigskip
\settabs \+ idd & haa & mn & second & deg & rm & arcsec & vstanda & stvsta & 
bstand & stbsta & rstand & strsta & istand & stista \cr
\hrule
\medskip
\+ ID & h & m  & s  & o  &  $'$  & $''$ &  $V$  & $\sigma V$  & $B-V$ & 
$\sigma_{B-V}$  & $V-R$  & $\sigma_{V-R}$ & $V-I$  & $\sigma_{V-I}$ \cr
\medskip
\hrule
\medskip
\+ 1  & 15 & 25 & 27.0  & 44 & 46 & 23.2  & 14.84 & 0.00  & 0.57  & 0.00  &
0.32  & 0.00  & 0.58  & 0.00 \cr
\+ 2  & 15 & 25 & 36.5  & 44 & 44 & 37.6  & 15.47   & 0.00  & 0.63  & 0.01  &
0.36  & 0.01  & 0.67  & 0.01  \cr
\+ 3  & 15 & 25 & 11.8  & 44 & 46  & 2.1  & 15.96   & 0.00  & 0.64  & 0.00  &
0.41  & 0.01  & 0.78  & 0.01 \cr
\+ 4  & 15 & 25 & 21.1  & 44 & 46 & 46.9  & 16.28   & 0.00  & 0.67  & 0.00  &
0.39  & 0.00  & 0.71  & 0.01  \cr
\+   5  & 15 & 25 & 38.6  & 44 & 43 & 20.3  & 16.72   & 0.00  & 0.60  & 0.01  &
0.32  & 0.01  & 0.63  & 0.01  \cr
\+   6  & 15 & 25 & 21.7  & 44 & 46 & 52.7  & 16.80   & 0.00  & 1.35  & 0.01  &
0.82  & 0.00  & 1.54  & 0.01  \cr
\+   7  & 15 & 25 & 38.8  & 44 & 44 & 30.1  & 17.65   & 0.06  & 1.29  & 0.07  &
0.83  & 0.08  & 1.34  & 0.08  \cr
\+   8  & 15 & 25 & 17.9  & 44 & 46 & 29.5  & 18.10   & 0.01  & 1.13  & 0.01  &
0.66  & 0.01  & 1.17  & 0.01  \cr
\+   9  & 15 & 25 & 39.1  & 44 & 44 & 46.1  & 18.48   & 0.05  & 1.32  & 0.07  &
0.67  & 0.08  & 1.25  & 0.07  \cr
\+  10  & 15 & 25 & 19.6  & 44 & 47 & 53.0  & 18.66   & 0.01  & 1.45  & 0.02  &
0.86  & 0.01  & 1.64  & 0.01  \cr
\+  11  & 15 & 25 & 19.3  & 44 & 45 & 58.1  & 18.72   & 0.01  & 0.47  & 0.01  &
0.04  & 0.02  & 0.53  & 0.02 \cr
\+  12  & 15 & 25 & 15.6  & 44 & 45  & 5.6  & 18.96   & 0.03  & 0.87  & 0.04  &
0.53  & 0.04  & 0.98  & 0.05  \cr
\+  13  & 15 & 25 & 32.7  & 44 & 44 & 29.9  & 19.01   & 0.01  & 0.53  & 0.02  &
0.37  & 0.01  & 0.69  & 0.02  \cr
\+  14  & 15 & 25 & 25.3  & 44 & 45 & 24.7  & 19.16   & 0.05  & 1.36  & 0.07  &
0.70  & 0.07  & 1.23  & 0.07  \cr
\+  15  & 15 & 25 & 16.4  & 44  & 47 & 28.4  & 19.65   & 0.02  & 1.39  & 0.04  &
0.88  & 0.02  & 1.71  & 0.03  \cr
\+  16  & 15 & 25 & 13.8  & 44 & 44 & 50.1  & 19.87   & 0.03  & 0.43  & 0.04  &
0.27  & 0.03  & 0.34  & 0.06  \cr
\+  17  & 15 & 25 & 26.6  & 44 & 43 & 55.3  & 20.14   & 0.03  & 1.06  & 0.06  &
0.80  & 0.03  & 1.23  & 0.04  \cr
\+  18  & 15 & 25 & 27.5  & 44 & 44 & 43.6  & 20.31   & 0.03  & 0.48  & 0.05  &
0.40  & 0.03  & 0.43  & 0.09  \cr
\medskip
\hrule

\vfill\eject

\noindent {\bf Table 3.} Statistical results of the photometric comparison. The
mean and standard deviations are based on N stars.
\bigskip
\settabs 4 \columns
\hrule
\medskip
\+ Range in magnitude & \hfill $ \Delta R $ \hfill & \hfill $ \Delta B \hfill $
 & \qquad $ \Delta (B-R) $ \cr
\+ & Mean $\pm \sigma $ \qquad N & Mean $\pm \sigma $ \qquad N &
Mean $ \pm \sigma $ \qquad N \cr
\medskip
\hrule
\medskip
\+ $ 14 < V < 17 $ & $ -0.02 \pm 0.16 $ \quad 6 & $ -0.04 \pm 0.19 $  \quad
6 &  $ -0.02 \pm 0.27 $ \quad 6 \cr
\+ $ 17 < V < 20 $ & $ +0.08 \pm 0.12 $ \quad 8 & $ -0.08 \pm 0.19 $  \quad
8 &  $ -0.16 \pm 0.20 $ \quad 8 \cr
\+ $ (B-R) < 1.4 $ & $ +0.02 \pm 0.15 $ \quad 9 & $ -0.02 \pm 0.14 $  \quad
9 &  $ -0.04 \pm 0.19 $ \quad 9 \cr
\+ $ (B-R) > 1.4 $ & $ +0.06 \pm 0.14 $ \quad 5 & $ -0.13 \pm 0.24 $  \quad
5 &  $ -0.26 \pm 0.25 $ \quad 9 \cr
\medskip
\hrule
\vskip 5 truecm
\centerline {\bf Caption to Figure}
\bigskip
\noindent {\bf Figure 1.} Comparison of present Photmetry with that of Monet
(1997). The difference in $B$, $R$ and $(B-R)$ in the sense (present--Monet) has
been plotted against present $V$ magnitude in Figure (a). Crosses, squares and 
open circles represent $\Delta B$, $\Delta R$, $\Delta(B-R)$ respectively. 
In Figure (b), $\Delta(B-R)$ has been plotted against present $(B-R)$. 
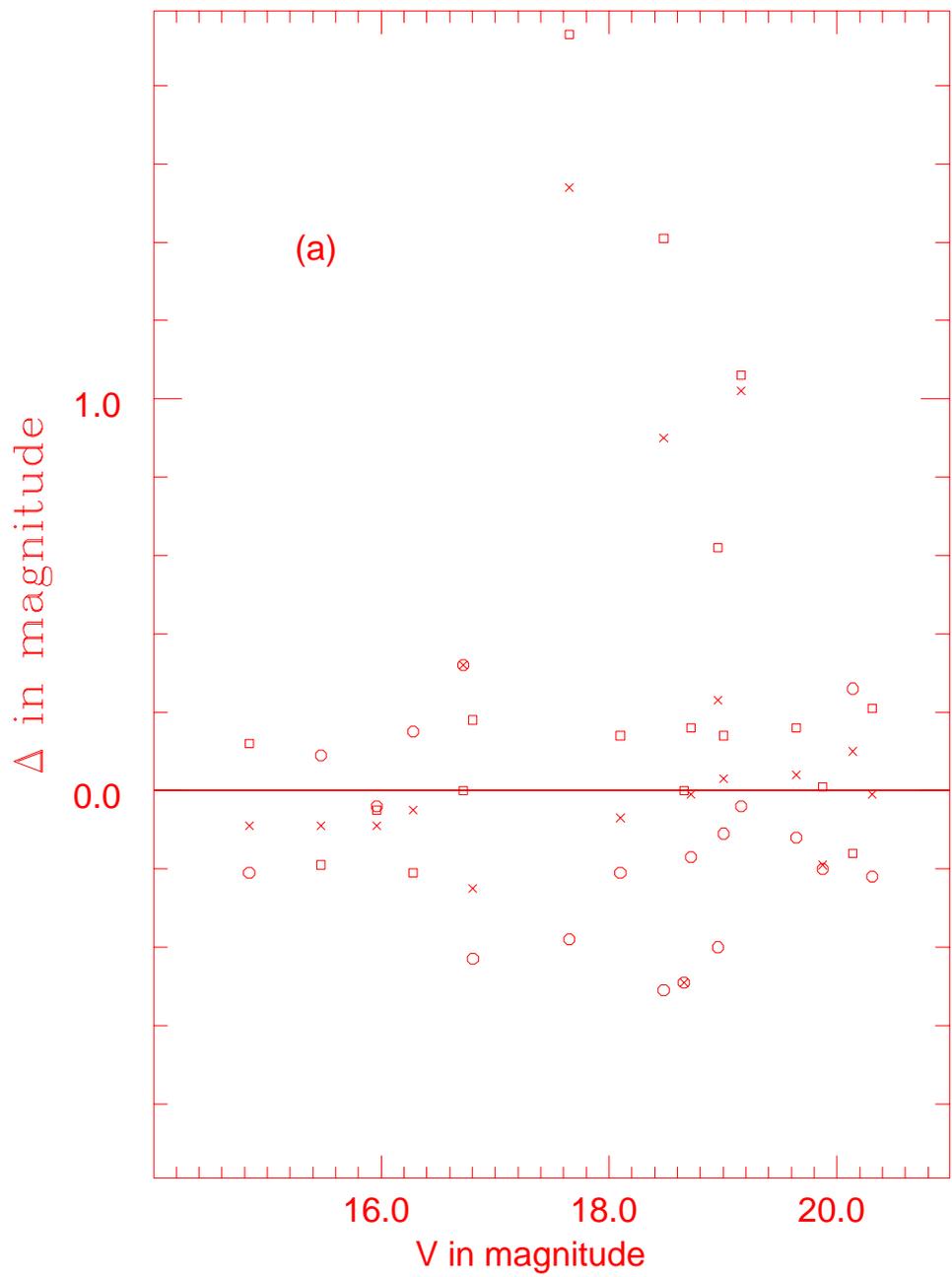
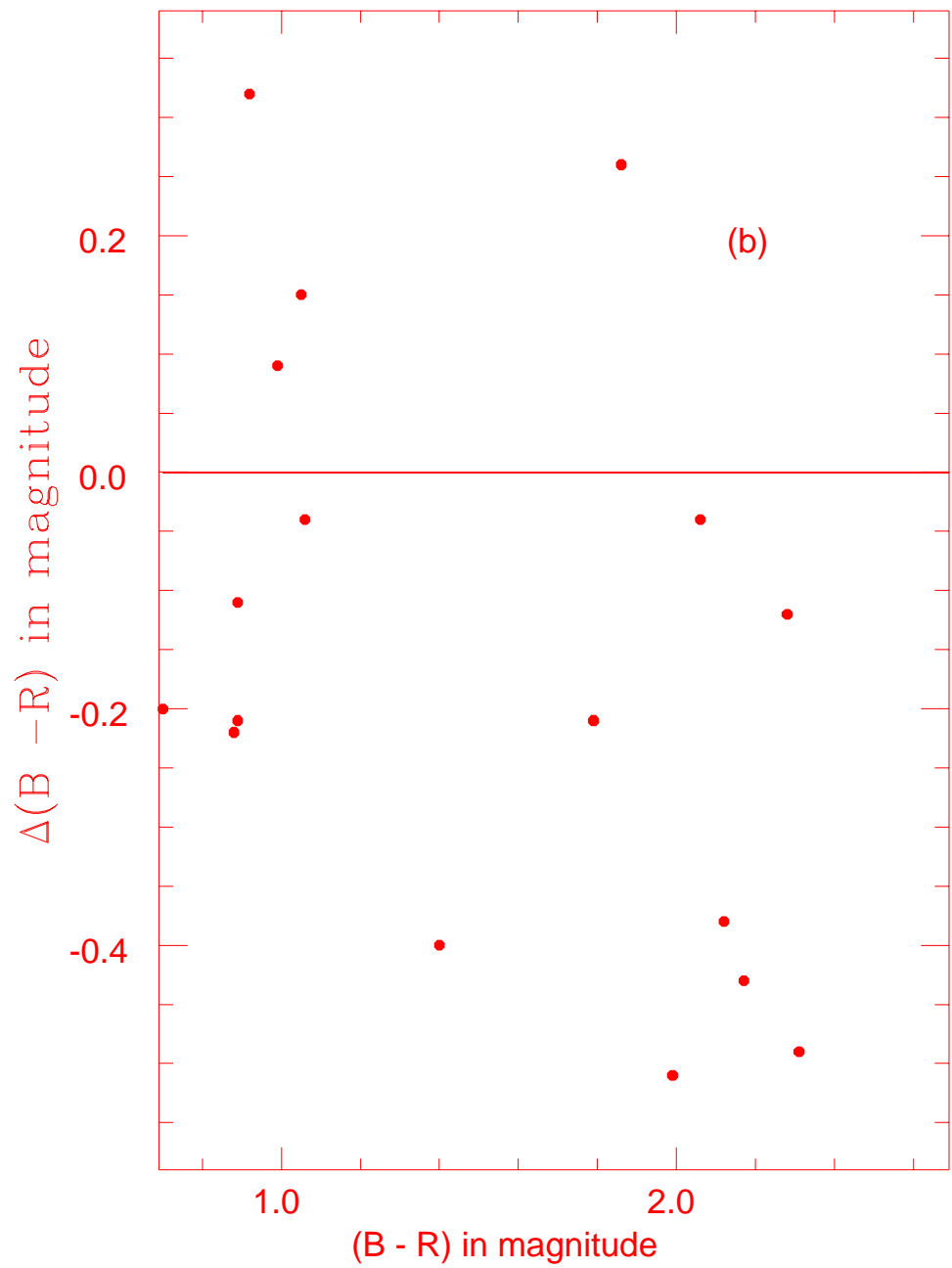
\bye